\documentclass[prl,groupedaddress,superscriptaddress,twocolumn,showpacs,amsmath,amssymb,a4paper]{revtex4-1}
\usepackage{amsmath}
\usepackage{graphicx}
\usepackage{graphics}
\usepackage{dcolumn}
\usepackage{bm}
\usepackage{color}
\usepackage{amsfonts}
\usepackage{amssymb}%
\usepackage{multirow}%

\definecolor{cyan}{rgb}{0.2,0.6,1}
\definecolor{red}{rgb}{1,0,0}
\definecolor{blue}{rgb}{0,0,1}
\definecolor{green}{rgb}{0,0.5,0}

\begin{document}

\title{Thermodynamics of animal locomotion}
\author{E. Herbert}
\thanks{These two authors contributed equally.}
\affiliation{Laboratoire Interdisciplinaire des Energies de Demain (LIED), CNRS UMR 8236, Universit\'e Paris Diderot, 5 Rue Thomas Mann, 75013 Paris, France}
\author{H. Ouerdane}
\thanks{These two authors contributed equally.}
\affiliation{Center for Energy Science and Technology, Skolkovo Institute of Science and Technology, 3 Nobel Street, Skolkovo, Moscow Region 121205, Russia}
\author{Ph. Lecoeur}
\affiliation{Center for Nanoscience and Nanotechnology (C2N), CNRS, Université Paris-Saclay, 91120 Palaiseau, France}
\author{V. Bels}
\affiliation{Institut de Syst\'ematique, Evolution, Biodiversit\'e, ISYEB, CNRS/MNHN/EPHE/UA UMR 7205, Mus\'eum national d'Histoire naturelle, Sorbonne Universit\'es, 45 rue Buffon, 75005 Paris, France}
\author{Ch. Goupil}
\affiliation{Laboratoire Interdisciplinaire des Energies de Demain (LIED), CNRS UMR 8236, Universit\'e Paris Diderot, 5 Rue Thomas Mann, 75013 Paris, France}

\date{\today}

\begin{abstract}Muscles are biological actuators extensively studied in the frame of Hill's classic empirical model as isolated biomechanical entities, which hardly applies to a living organism subjected to physiological and environmental constraints. Here we elucidate the overarching principle of a \emph{living} muscle action for locomotion, considering it from the thermodynamic viewpoint as an assembly of actuators (muscle units) connected in parallel, operating via  chemical-to-mechanical energy conversion under mixed (potential and flux) boundary conditions. Introducing the energy cost of effort, $COE_-$, as the generalization of the well-known oxygen cost of transport, $COT$, in the frame of our compact locally linear non-equilibrium thermodynamics model, we analyze oxygen consumption measurement data from a documented experiment on energy cost management and optimization by horses moving at three different gaits. Horses adapt to a particular gait by mobilizing a nearly constant number of muscle units minimizing waste production per unit distance covered; this number significantly changes during transition between gaits. The mechanical function of the animal is therefore determined both by its own thermodynamic characteristics and by the metabolic operating point of the locomotor system. 
\end{abstract}

\maketitle

\paragraph*{Introduction}
The ability to move is a fundamental characteristic of animal life \cite{Nathan2008}, the study of which from a physical viewpoint dates back to Aristotle \cite{Aristotle}. Whether in the air, under water or on the ground, animal locomotion in its rich variety of modes and purposes, rests on the active association of three of the main systems that compose the animal body: the skeleton, nervous system, and muscles \cite{Dickinson2000,Ijspeert2003,Biewener2018}. By active association, we mean that to set the whole body or part of it in motion, the somatic nervous system sends control signals that trigger chemical reactions in the skeletal muscles, which in turn act mechanically on the bones. Notwithstanding the rather detailed understanding of some essential aspects of animal locomotion, a complete holistic physics description of its mechanisms, including the couplings between the body actors (nerves, muscles, bones) and boundary conditions (environment), is yet to be achieved: outstanding questions concerning, e.g., neuromuscular control, notably considering overload and fatigue problems, biomechanics and sex-specific patterns, and energetics to name just a few, remain to be addressed \cite{VanLeeuwen1999, Nelson2016, NeuroErgo2018, Bruton2013, Munoz-Martel2019, Palmeira2014, Lewenstam2015}.

Physiological properties of living organisms such as, e.g., temperature, pressure, and chemical species concentrations in fluids, which can be described as thermodynamic variables, are maintained within certain ranges by homeostatic mechanisms to ensure steady-state internal working conditions \cite{Cannon1963}. Further, since the thermodynamic description of the energy conversion process permitting muscle motion does not require consideration of all the intricate biochemical processes at the heart of the complex body's regulatory system, Onsager's close-to-equilibrium force-flux formalism \cite{Onsager1931} is very well suited for the study of metabolism under muscle load. In a recent work, we developed such a nonequilibrium thermodynamics model to understand the chemical-to-mechanical energy conversion process under muscular effort \cite{Goupil2019}, considering living organisms as open thermodynamic systems that exchange energy and matter with their environment. We derived Hill's widely used empirical muscle equation \cite {Hill1938} from the principles of thermodynamics, provided the description of the response of the muscle in terms of active impedance, and critically discussed the so-called maximum power principle \cite{Odum1995}, which was formulated after Lotka's theory of energy optimal consumption based on the energy/efficiency trade-off and exergy \cite{Lotka1922b}. We also showed that for a generic energy conversion engine, living or not, power maximization \cite{Lotka1922a,Lotka1922b}, entropy minimization \cite{Jaynes1980}, efficiency maximization, or waste minimization states are only specific operation modalities \cite{Goupil2019,Goupil2020}. 

Animals manage their energy expenditure as their movement is constrained both by needs and availability of metabolic resources. Broadly speaking, the most efficient systems minimize energy dissipation and entropy production at the cost of being also the most constrained in terms of working conditions and use, while systems that do not boast high efficiency or power, may have a wider range of use and therefore marked robustness. A key question to consider is the existence of an energy (basal) flow at rest, to deduce the constraints due to energy conversion specific to living systems \cite{Goupil2019,Goupil2020}. Further, as animals may change gaits or, more generally, their locomotor behavior, the questions of energy efficiency and oxygen consumption variations on the one hand, and of the specific muscular mechanisms permitting transition, are yet to be precisely answered. Indeed, though Hill's muscle model \cite {Hill1938} is extensively used in biomechanics, it is important to remind the reader that Hill's studies of muscles were performed using dissected muscles extracted from dead animals, while for energy efficiency and oxygen consumption economy, knowledge of the actual oxygen cost of transport (COT) is needed, meaning the study of a living muscle and its boundary conditions.

In this Letter, we address the generic question of thermodynamic constraints applied to an animal, assessing their impact on the animal's effort production, using instantaneous oxygen consumption experimental data. Building on \cite{Goupil2019,Goupil2020}, we develop an integrated framework for animal locomotion, which may also bridge calorimetric measurements of muscles \cite{Kaiyala2011}, the dynamic energy budget phenomenological approach based on balance and conservation laws \cite{Kooijman2000, Sousa2006}, and biological studies based on the analysis of proxies such as oxygen consumption. We thus establish: (i) the link between oxygen consumption during muscular effort of moderate intensity and our thermodynamic formalism for metabolism \cite{Goupil2019}, to use oxygen consumption to characterize and compare the modes of movement \cite{Tucker1975} and as a proxy of the output flux of low-grade waste energy; (ii) a basic model describing a complex organism as an association of muscle fibers, in light of which we discuss experimental data \cite{Hoyt1981}, providing insights from which we can draw general conclusions on animal locomotion from the energetic viewpoint. 

\paragraph*{Animal activity and oxygen COT}
\label{section2} 
An animal has to arbitrate between several operating points, depending on the desired optimization, say, e.g., maximum efficiency, maximum power production, or minimum waste flow per unit of power produced. The constitutive metabolic force ($F_{M}$)-intensity ($I_{M}$) equations describing an organism's overall energy balance, considering an assembly of $N$ (identical) muscle units \cite{SuppMat}, producing the mechanical power $P_M$ connected in parallel to the chemical energy source and sink, read \cite{Goupil2019,SuppMat}: 
\begin{eqnarray}
\Phi_{+} &=& N \varphi_+ = \alpha \mu_{M+} I_{M} + \Delta \mu_M/R_E \label{eq:PHI+} \\
\Phi_{-} &=& N \varphi_- = \alpha \mu_{M-} I_{M} + R_{M}I_M^2 + \Delta \mu_M/R_E \label{eq:PHI-} \\
P_{M}    &=& N p_M = F_{M} I_{M} = \Phi_{+} - \Phi_{-} \label{P}
\end{eqnarray}
where $\Phi_{\pm}$ are the incoming from the source/outgoing to the sink energy fluxes, $\Delta\mu_M=\mu_{M+}-\mu_{M-}$ is the chemical potential difference across the conversion zone, which has efficiency $\eta=P_M/\Phi_+$, and the resistances $R_M$ and $R_E$ characterize dissipation and entropy production due to the coupled ($\alpha$) transport of energy and matter. Note that the zero metabolic intensity configuration $I_M = 0$ describes the organism globally at rest with a \emph{nonzero} basal residual energy consumption $B \equiv Nb\approx \Delta \mu_{M}/R_E$ \cite{SuppMat}. The quantities $\varphi_{\pm}$, $i_M$, $p_M$, and $b$ are defined as per muscle unit. For aerobic efforts, i.e., of limited duration and intensity, that the respiratory chain is involved in at the end of the Krebs cycle, via cytochrome oxidases, shows that the measured oxygen consumption is a proxy of the power fraction rejected $\Phi_{-}$ \cite{Castresana1994}. One can then define the energy cost of effort index $COE_-$ as a measure of the energy required per unit of muscular effort, i.e. $COE_- \equiv \Phi_ {-}/I_M$, which in the framework of \cite{Goupil2019,Goupil2020}, reads:
\begin{equation}\label{eq:COE}
COE_- = a_0+ R_{M} I_{M} + \Delta \mu_{M}/(R_E I_M) 
\end{equation}
\noindent with $a_0=\alpha \mu_{M-}$ being Hill's constant parameter \cite{Goupil2019}. The term $\Delta \mu_{M}/(R_E I_M)\approx B/I_M$ is only dominant in situations where the metabolic intensity $I_M$ is low.
\begin{figure}[ht]
	\includegraphics[width=0.3\textwidth]{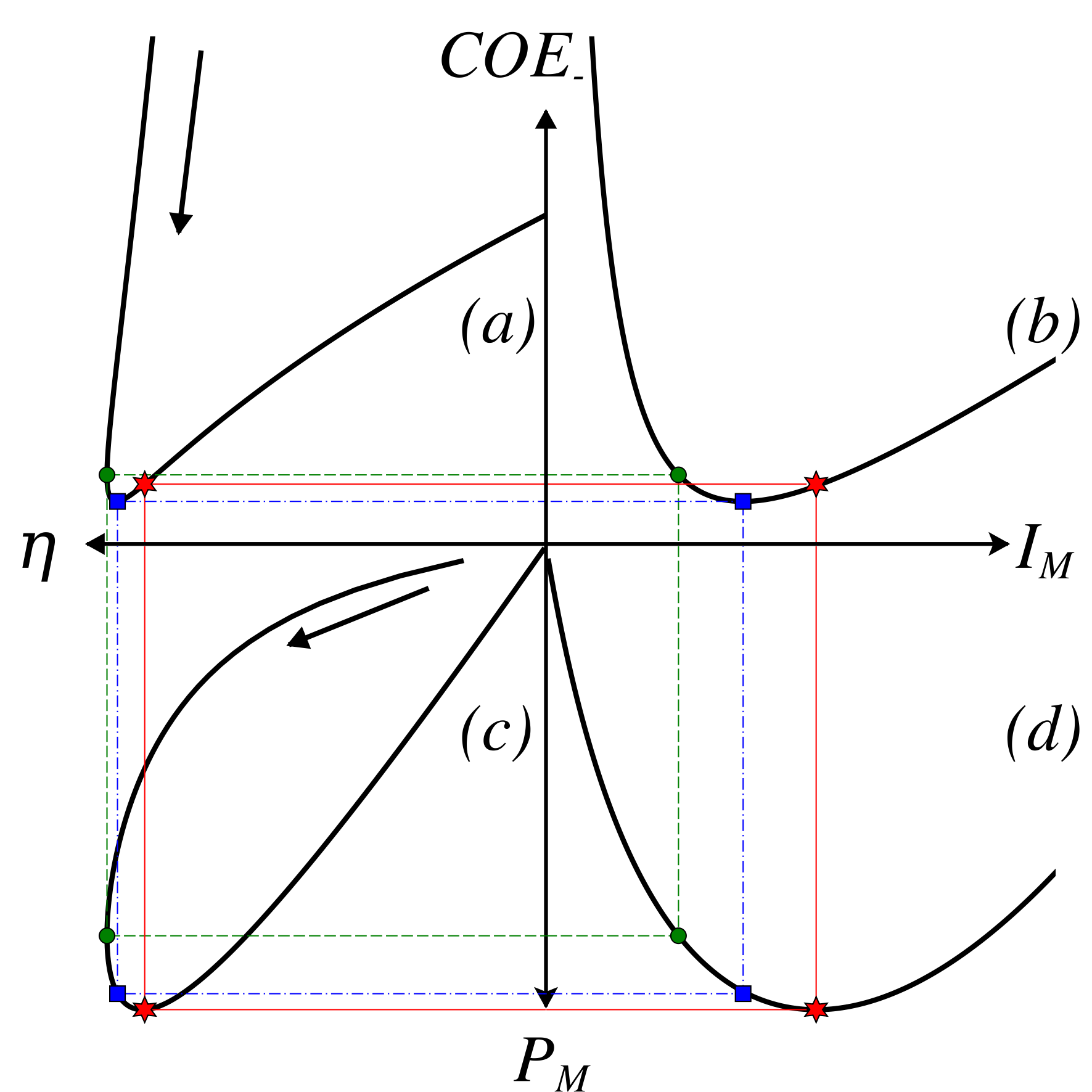}
	\caption{Four-quadrant plot of $COE_-$ (North direction), $\eta$ (West), $I_{M}$ (East) and $P$ (South): $(a)$ $COE_-$ vs $\eta$, $(b)$ $COE_-$ vs $I_{M}$, $(c)$ $P_M$ vs $\eta$, $(d)$ $P_M$ vs $I_{M}$, from Eqs.~(\ref{eq:PHI+}-\ref{eq:COE}), using $\Phi_+ = \frac{\mu_+ - \mu_{M+}}{R_+}$, with $R_M=10$, $R_+=10$, $R_-=0$, $R_E = 100$, $\alpha=0.5$, $\mu_- = \mu_{M-}=10^{-4}$ and $\mu_{+}=100$ (all in arbitrary units). The arrows show the direction along which $I_{M}$ increases. The red star symbol (resp. blue squares and green dots) indicates the position of $P_{\max}$ (resp. $COE_-^{\ast}\equiv\min(COE_{-})$ and $\eta_{\max}$).}
	\label{QuatreQuadrant1}
\end{figure}

The relevant quantities for the energetic description of an animal's muscular activities are summarized in the reduced set: $\left\{P;\eta; COE_-;I_{M}\right\}$, which can be put together in a single four-quadrant chart as shown in Fig.~\ref{QuatreQuadrant1}. The overall observed behavior resembles that of a thermodynamic system with the two usual optimum working points, namely the maximum efficiency $\eta_{\max}$ and the maximum power $P_{\max}$, which can be readily identified. However, a third optimum working point is also evidenced here, which corresponds to the minimization of the waste flux $\Phi_{-}$ per unit of metabolic intensity $I_{M}$, leading to a minimal value for $COE_-$, denoted $COE_-^{\ast}$. These three optima correspond to three different values of the metabolic intensity $I_M$ shown in Fig. \ref{QuatreQuadrant1}-(c): the organism first sees both its efficiency and power increase with $I_M$ before the points of maximum efficiency, minimum $COE_-$, and finally maximum power are \emph{successively} reached. Beyond the latter point, the organism is in a physiologically unfavorable situation, which can, at the extreme, lead to exhaustion. 

The oxygen COT, quantifies the total amount of energetic waste by unit mass of animal to perform a unitary displacement; it is routinely used for categorizing animal species with respect to their motion efficiency \cite{Tucker1975,Wickler2000,VanDenThillart2004,Chappell2009,Scantlebury2014,Williams2014}. We now write $COT~\equiv \Phi_{-}/v$, with $v$ being the animal's velocity, assuming a linear relationship $I_{M}=kv$ (with $k>0$) between the metabolic intensity and the animal displacement velocity \cite{SuppMat}. The mechanical power $P_M$ produced by the organism is necessarily equal or higher than the power $P_r$ required to enable the displacement under various experimental conditions: ascent or descent, headwind or backwind, load carried or assistance with the motion; hence the number of units involved $N$ and the metabolic intensity $i_M$ increase with the required mechanical power, which drives the growth of the metabolic power. Here, assuming that $N$ varies linearly with $i_M$ \cite{SuppMat} and hence with $v$, we get: 
\begin{equation}\label{eq:COT}
	COT = \frac{N}{N_H} \left(a_0 k + R_M k^2 v + \frac{B}{v}\right) = k\frac{N}{N_H} COE_-
\end{equation}
\noindent from Eq.~\eqref{eq:COE}, with $N_H$ being the maximum (fixed) number of muscle units that can be mobilized for an effort, as for a Hill-type of effort \cite{Hill1938,Goupil2019}. The system's response thus is characterized by only three parameters: $a_0 k$, $R_{M} k^2 $ and $B$, the latter two being dependent on $N$, unlike the former, which is directly related to the ``extra heat'' term in Hill's equation \cite{Goupil2019,SuppMat}. The speed for which the $COT$ is minimum, is $v \equiv v^{\ast} = \frac{N}{k} \sqrt{\frac {b}{r_ {M}}}$, from which we get the $COT$ minimum value for a fixed $N$:
\begin{equation}\label{eq:cot*}
	COT^{\ast} = a_0 k + 2 k \sqrt{r_M b}
\end{equation}
which remarkably turns out to be independent of the number $N$ of muscle units put in action during the effort. $COT^{\ast}$ is therefore an \emph{intrinsic} characteristic of the organism, independent of the imposed experimental configuration, and $B_v=\frac{N}{N_H}B$, $R_v=\frac{N}{N_H}R_M$ and $a_v=\frac{N}{N_H}a_0$ thus become experimentally directly accessible. 

\paragraph*{Gaits modeling; the illustrative case of horse motion}
\label{sec:application}
We now focus on the documented case of horse motion studied by Hoyt and Taylor \cite{Hoyt1981}. Tucker showed that the oxygen minimal COT strongly depends on the body mass of animals regardless of shape and habitat (water, terrestrial and arboreal surfaces, and in air), and for various types of locomotor modes and mechanisms \cite{Tucker1975}. The data are confirmed by studies of quadrupedal, bipedal, flying and swimming animals \cite{Tucker1970,Nishii2000,Nakatsukasa2004}. The horse is one classical, illustrative animal of studies in energetics of locomotion \cite{McMiken1983,Jones2006,Gerard2013}. Three main modes of displacement are usually adopted by a horse: walk, trot and gallop, each gait representing different biomechanical working conditions. We therefore selected this one particular species for our study, from which general conclusions on animal locomotion can be drawn. 

\begin{figure*}
\begin{tabular}{cc}
\includegraphics[width=0.35\textwidth]{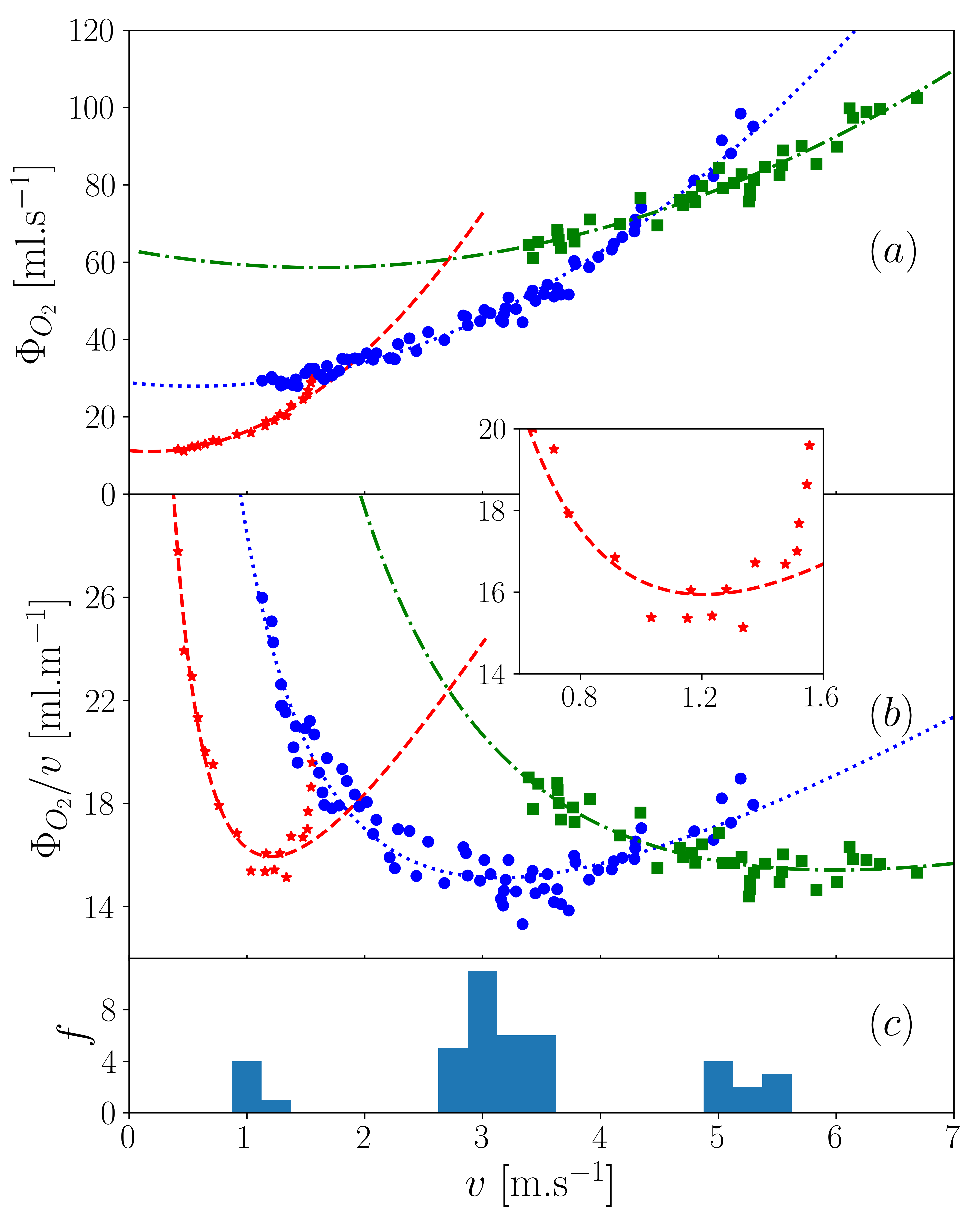}  &
\includegraphics[width=0.35\textwidth]{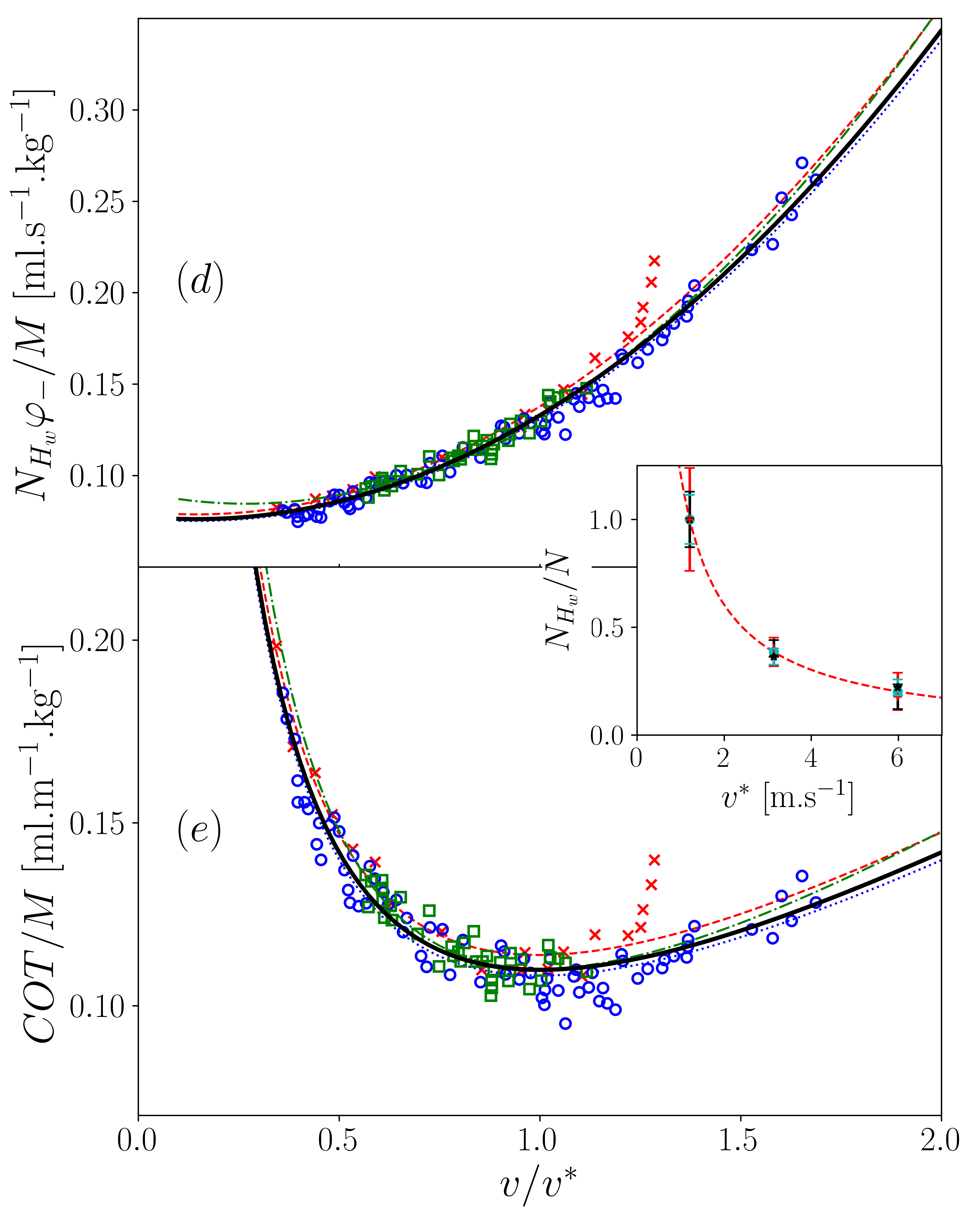} 
\end{tabular}
\caption{On the left panels, experimental data from \cite{Hoyt1981} - horse B of mass $M=140$ kg: oxygen flux $\Phi_{O_{2}}$ and $COT \equiv \Phi_{O_ {2}}/v$, plotted against the speed $v$ for walk (red stars), trot (blue dots), and gallop (green squares), and their fits with our modeling, Eq.~\eqref{eq:COT}, and fitting parameters in Table \ref{tab:bilan_thermo}. Note the $COT$ dramatic increase for the high-speed walk shown in the inset: this slope change marks the change of muscular effort regime in this region \cite{SuppMat}. Figure~\ref{fig:Hoyt}-c shows for each gait the histogram  of the observed range of velocities naturally used by the horse let free to run on the ground.  
On the right panels, $N_{H_w}\varphi_{-}/M$ and the specific $COT/M$, are plotted against the scaled speed $v/v^{*} \propto i_M$. In both cases, the thick dark line is a two-parameter fit of the aggregated data. All gaits are considered as a collection of a different number of activated muscle units; the ratio ${N_{H_w}}/{N}$ vs $v^{*}$ is shown in the inset. The ratio ${v_w^{*}}/{v^{*}}$ is represented with blue dots, ${B_{v_w}}/{B_{v}}$ with red $+$, and ${R_ {v}}/{R_{v_w}}$ with dark stars. The red dotted line $1/v^{*}$ serves as a guide for the eyes for comparison purposes.}
\label{fig:Hoyt}
\end{figure*}

The measurement data for each of the considered gaits are reported in Fig.~\ref{fig:Hoyt}-$a$. Walk is chosen as the reference gait and the related quantities are all denoted with the subscript $w$, e.g., $N_H\equiv N_{H_w}$ for the walk. When the animal was let free to run on the ground, some ranges of speeds were naturally never used by the horse, for any sustained period, as shown in the histograms Fig.~\ref{fig:Hoyt}-$c$. Whichever gait was adopted, the speeds chosen by the animal were systematically close to the speed corresponding to \emph{minimal} $COT$, i.e. close to the  point of minimal waste rejection per unit of displacement. Increasing the animal motion velocity while maintaining a constant metabolic intensity per fiber requires increasing the number of muscle units: this is achieved only by a change of gait. Interestingly, this may be related to studies of robots mimicking walking bipeds or quadrupeds \cite{Semini2011,Gehring2013,Hutter2013,Seok2015,Yesilevskiy2018,Wu2019}, though for robots, energy optimization is based on a trade-off between the number of limbs for motion and the energy consumption of their motorization, as their actuators do not have a number of subunits, which could be activated on demand \cite{Yang2011,Vanderborght2013}. 
\begin{table}[th]
\centering
\begin{tabular}{l||c|c|c|c|c}
\hline
& $a_v k$ 		   & $R_v k^2$       & $B_v$            & $v^{\ast}$ 	  & ${COT}^{\ast}$\\
& [N$\cdot$kg$^{-1}$]& [s$^{-1}$] & [W$\cdot$kg$^{-1}$]& [m$\cdot$s$^{-1}$]& [N$\cdot$kg$^{-1}$]\\[3pt] 
\hline
\rule{0pt}{4ex}W	& $-0.37\pm 0.34$   & 1.10$\pm$0.18   & 1.60$\pm$0.13    & 1.21$\pm 0.11$  & 2.27$\pm$0.43\\
~&~&~&~&~&~\\
T	& $-0.46\pm 0.17$   & 0.42$\pm$0.03   & 4.11$\pm$0.20    & $3.14 \pm 0.14$ & 2.16$\pm$0.21\\
~&~&~&~&~&~\\
G	& $-0.80\pm 0.78$   & 0.24$\pm$0.08   & 9.01$\pm$1.71    & $5.99 \pm 1.61$ & 2.20$\pm$0.98\\[3pt]
\hline
		\end{tabular}
\caption{Horse thermodynamic characteristics. The fitting parameters for walk (W), trot (T) and gallop (G), are obtained from the experimental data of \cite{Hoyt1981} with Eq.~\eqref{eq:COT} and using the conversion factor 20 J$\cdot$ml$^{-1}$ O$_{2}$ for the heat produced as oxygen is consumed during the effort \cite{thornton1917}.} 
		\label{tab:bilan_thermo}%
\end{table}
The fitting curves obtained with Eqs.~\eqref{eq:PHI-} and \eqref{eq:COT} are in remarkable qualitative agreement with all the experimental data of Hoyt and Taylor \cite{Hoyt1981} as shown in Fig.~\ref{fig:Hoyt}. As the metabolic intensity increases, the number of muscle units involved in the motion is likely not conserved, both within a gait and between gaits; so it is important to determine whether the intragait variation remains small or not compared to the intergait variation: if the former is small, one can then assume that the fitting parameters within a same gait can be taken as constants. The oxygen flux fitting curve $\Phi_{\rm O_2}$ in Fig.~\ref{fig:Hoyt}, in good quantitative agreement with the experimental data, is simply a polynomial of degree 2; this justifies in what follows the use of constant fitting parameters, thus neglecting higher-order contributions \cite{SuppMat}. 

As $COT^{\ast}$ is a constant independent of the number of muscle units involved, the law governing the modulation of the number of muscle units $N$ remains the same for \emph{all} gaits. When the number of fibers is fixed, $a_0$ is expected to be the same for all gaits; varying the number of fibers within the same gait should lower the value of $a_0$. We observe that the numerical value $a_v $ is found to be slightly negative but essentially the same for all gaits. It is therefore legitimate to consider that (i) the variation in the number of intragait units remains moderate, i.e., of the order of 10 \% \cite{SuppMat}, though we cannot quantify it more accurately; (ii) this variation is identical for each gait. We may then safely assume that $a_v \approx a_0$, $ R_v \approx R_M $, and $ B_v \approx B$. We can also consider that Eq.~\eqref{eq:cot*}, giving $COT^*$ at a constant $N$, is a quite accurate approximation, with fitting parameters  $R_{v}$ and $B_v$, also assumed constant within the same gait, linked to the intrinsic parameters $r_{M}$ and $b$. One can find, in particular, that $k\sqrt{R_{v}B_v} = k\sqrt{r_{M}b}\approx 9 \pm 0.5$ ml$\cdot$m$^{-1}$ is indeed constant. 

If we now consider the intergaits behavior, we find as expected that the resistance $R_v\approx r_M/N$ decreases when the gait increases, as in this case a growing number of muscle units are put to work. From a thermodynamic viewpoint, this amounts to increasing the number of thermodynamic engines operating in parallel, rather than increasing the intensity of operation of each of them. As a result, the unit metabolic intensity is not unduly increased, thus limiting the influence of the quadratic dissipative terms, and the multiplication of the units put in parallel leads to a basal power value $B_v\approx Nb$ increase by the same multiplicative factor. We find that $B_v$ is approximately increased by a factor of 2.5 from walk to trot, and by 5.6 from walk to gallop. Therefore, the variation of $N$ between two gaits is found to be significantly greater than it is within the same gait, which allows us to safely assume that the number of fibers is constant for a given gait, and define the scaled velocity: 
\begin{equation}\label{eq:v*}
	V = v/v^{\ast}\approx \sqrt{r_M/b} \times i_M
\end{equation}
\noindent which establishes a linear relationship between $V$ and $i_{M}$, the proportionality coefficient $\sqrt{r_{M}/b}$ being entirely determined by the metabolism of a single muscle unit. 

Let us now evaluate more precisely the number of muscle units involved during displacement. The actual number $N_{H_w}$ of muscle units involved for walk is of course not known in the experiment. However, from Eq.~\eqref{eq:COT}, one can derive the relative number of muscle units, $N/N_{H_w}$, put in action in the two other gaits: $N_{H_w}/N=v_w^{\ast}/v^{\ast}$, with $N_{H_w}\sqrt{b/r_{M}}=v_w^{\ast}/v^{\ast}\sqrt{B_{v}/R_{v}}$. The parameters $N$, $v^{\ast}$, $B_{v}$ and $R_{v}$ are connected through the identities: $
N_{H_w}/N=v_w^{\ast}/v^{\ast}=R_{v}/R_{v_w}=B_{v_w}/B_{v}$. The ratio $N_{H_w}/N$ is shown in the inset of Fig.~\ref{fig:Hoyt}-$e$ as a function of $v^*$. For a given metabolic intensity $i_{M}$, the horse can increase its speed when changing its gait from walk to trot and from trot to gallop, mainly by increasing the number of muscle units put in action, by a factor $N/N_{H_w}$, respectively $2.59\pm 0.04$ and $4.95\pm 0.06$. From Eq.~\eqref{eq:COE}, we also recover the waste energy output flux $N_{H_w}\varphi_{-} = (v_w^{\ast} v/v^{\ast})\times COT$, resulting from the oxygen consumption of $N_{H_w}$ fiber bundles for different gaits, shown in Fig.~\ref{fig:Hoyt}-$e$. As expected, all curves collapse into a \emph{master} curve, and including the scaled velocity $V$ in Eq.~\eqref{eq:COT} finally yields $\widetilde{COT}$, which depends on only two adjustable factors:
\begin{equation}\label{eq:cot_tilde}
	\widetilde{COT}=a_0 k + k\sqrt{r_M b} \left(V+1/V\right)
\end{equation}
\noindent The dark thick line of Fig.~\ref{fig:Hoyt} represents the fitting curve, Eq.~\eqref{eq:cot_tilde}, for the all aggregated data from which we extract: $a_0 k =-0.37\pm 0.08$~N$\cdot$kg$^{-1}$ and $k\sqrt{r_M  b}=1.284\pm 0.004$~N$\cdot$kg$^{-1}$. Each $COT$ curve can be described using a minimal set of three parameters: $B$, $R_{M}$ and $a_0$. Note that $a_0 k$ is found to be slightly negative as a result of the modulation of the number of fibers involved in the displacement within the same gait, but as the feedback resistance $R_{\rm fb}\propto a_0$ \cite{Goupil2019}, feedback therefore appears as a positive contribution to the available power $P_M$. 

\paragraph{Conclusion}
\label{sec:conclusion}
An animal left free to choose its locomotion speed operates at minimum waste production per unit distance covered. A muscle may be considered as a system composed of muscle units connected in parallel, the number of which in action varies little in time for the same gait, but substantially changes during transition to a different gait, thus showing from a thermodynamic viewpoint how an animal's muscles operate in concert to sustain a particular effort \cite{Perry1988}. The master curve of the number of muscle units put in action clearly confirms this result. Our model provides a valuable means to test the motor behaviors resulting from muscular actuators among living organisms, e.g., prey chasing, courtship display, using a reduced set of physiological parameters, easily extractable from the literature or from experiments, within a comparative and evolutionary framework. Further, one may envisage to adapt it to the cardiac muscles, which, like skeletal muscles are made of striated tissues, though they differ greatly in purpose and operation mode \cite{SuppMat,Adams1980}. Weaving a conceptual link between the skeletal muscles and the actuators used in soft robotics \cite{AlbuSchaffer2008,Kim2013,Ijspeert2020}, our model may be used to bridge studies of animal locomotion and robot locomotion in terms of COT and gait adaptation \cite{Christensen2010,Larsen2011,Kashiri2018,Esfanabadi2019}, as adaptability, acquired by an increase of the number of limbs, competes with the need to minimize energy consumption. An adopted solution cannot be simultaneously adapted \emph{and} adaptable: the more efficient the solution, the narrower the optimal operating range, implying that optimization in the sense of adaptability to changing environments and, on the contrary, adaptability to a stable environment, results in differing evolutionary strategies \cite{Goupil2020}.

\begin{widetext}

\section*{Supplemental Material}

\subsection{Definition of the locomotor system's muscle unit and assumptions}
Each skeletal muscle may be seen as a system that has two separate points of attachment to a bone (tendons) to which it is connected. This system, made of striated tissue, can be said to be constituted of fibers regularly organized, connected in parallel along the muscle from one tendon to the opposite one. In our work we are interested in the thermodynamics of the locomotor system, rather than the animal's biomechanics. The number of muscles in an animal body can reach several hundreds, making the body a complex system from the biomechanics viewpoint, but from the structural viewpoint, as a first approximation in our thermodynamic model, we reduce the animal body to its locomotor system, which brings us to consider it as a ``parallel circuit'' \emph{for motion}, and whose components, the skeletal muscles, are bundles of fibers organized in a parallel configuration. In some sense, we separate muscles according to their functional specialization into two systems, the locomotor one, and the structural or posture one.

From the intuitive mechanistic viewpoint, each muscle unit may be viewed as a spring; the stretching and contraction of the muscle corresponding to a given developed force by the said muscle, depend on the number of springs in action associated in parallel and not in series. In our work we consider the muscles that produce the mechanical force responsible for the motion, i.e. the limb muscles, while the other muscles responsible for the stability of the structure, i.e. the bearing of the body while in motion, are considered as part of the said structure rather than the locomotor system. These latter muscles are of the ``slow type'': they shorten slowly and involve low power, thus consuming a low amount of energy \cite{He2000}; this implies that their contribution to the energetic budget may be included in the basal. It is a simplifying assumption whereby the muscles we model are essentially the actuators in charge of mechanical power production for motion, and it is quite reasonable to assume that these particular muscles act in parallel as a given number of them contribute in concert to the total force that results in the desired motion, while the other type of muscles even when contracted do not produce power for motion, though they are necessary to participate in the overall balance of the structure in motion. A given gait requires a number of muscle units working in parallel in a muscle; in the parallel configuration, all the units need not be mobilized: the number depends on the mechanical force required for the locomotion at a given gait. 

Further, from the thermodynamic viewpoint, the physical power is determined by the product of the potential difference and the flow. The conversion from the metabolic chemical power to mechanical power is of course based on the chemical potential difference involved in the mechanisms. It is clear that this chemical potential difference in a fiber cannot be increased at will as for a system that would be driven by electrochemical battery cells in series, the number of which could be increased. On the other hand, the flow of ATP reagents is scalable. We can therefore see that intrinsically the adaptation of the metabolic power in response to a particular locomotion need is based on the mobilization in parallel of the mechanical power production units. The control variable in the metabolic chemical-to-mechanical energy conversion is the flux, not the potential difference.

It is also of interest to note that like the skeletal muscles, the cardiac muscles are made of striated tissue, although they differ greatly in purpose and operation mode. The former, to set a limb or the whole body in motion, act as a result of a voluntary, conscious stimulus (somatic system), while the latter, which act as a pump for blood circulation, depend on the autonomic nervous system, implying involuntary contractions without the need for a conscious stimulus. This difference entails differing physiological mechanisms that drive the energy fluxes \cite{Adams1980}. It is possible to envisage a thermodynamic model of the cardiac muscle and its operating points as long as the incoming and outgoing energy flows are properly characterized. The output power can be calibrated from the blood flow rates obtained. The metabolic intensity can be obtained from the heart rate. The sensitive point is the basal power which should be gauged from the resting heart rate, which may calibrate the point of effort and zero metabolic intensity.

\subsection{Recap of the thermodynamics of metabolism \cite{Goupil2019}}
For simplicity, we assume that the muscle units produce an effort of limited duration, which prevents any saturation effect due to the presence of waste, including secondary metabolites production. In our approach, a living muscle or even a complete organism, is a system composed of a source and a sink, both connected to a locus where energy conversion actually occurs as depicted in Fig.~\ref{figure1}. The coupled transport of energy and mass fluxes through the conversion zone is characterized by the resistance $r_E$, associated to the energy flux, and the resistance $r_M$, associated to the mass flux, thus yielding dissipation and entropy production. The source reservoir (at potential $\mu_+$) includes the resource, in the form of chemical energy, and the sink (at potential $\mu_-$) is the receiving zone for energetic, chemical and thermal wastes, rejected after completion of the conversion process. Two resistive dipoles, which ensure the connection of the conversion zone to both reservoirs, define the boundary conditions for the access to the resource with resistance $r_+$, and the waste rejection with resistance $r_-$. The construction of our model thus requires solely the chemical potential as a thermodynamic potential, which is perfectly justified inasmuch the chemical potential is a physical quantity that can be absolutely measured. As for other thermodynamic systems, the modification of the boundary conditions \cite{Ouerdane2015} generate feedback loops \cite{Goupil2016} that largely govern the overall behavior of the system. Note that our approach is quite similar to that for thermoelectricity where heat (microscopic-scale energy) is directly converted into electric -- usable, at the macroscale -- work, and for which performance is related to the working conditions imposed by the boundary conditions  \cite{Apertet2014}. 
\begin{figure}[ht] 
\includegraphics[width=0.5\textwidth]{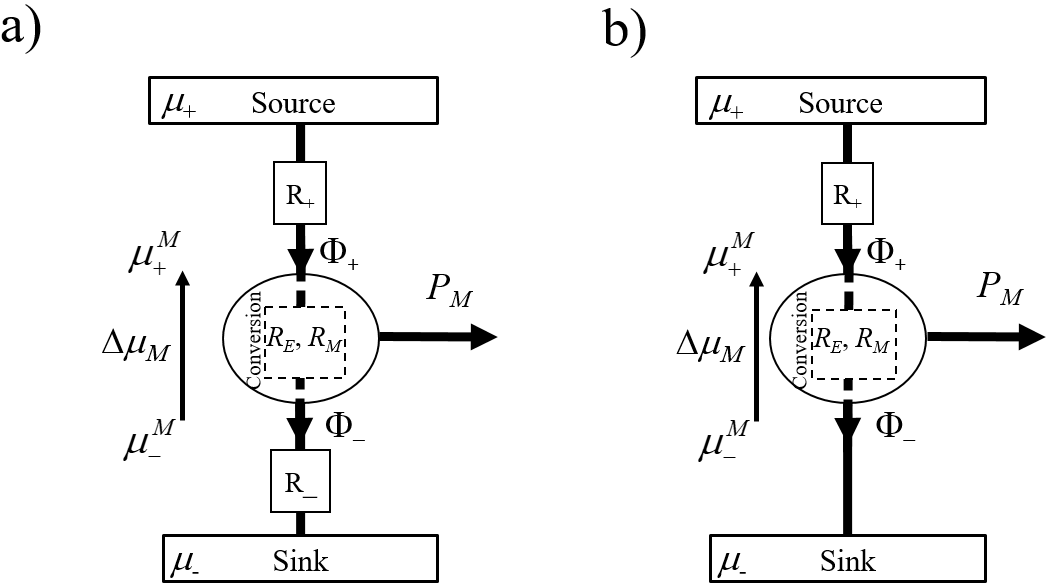}
\caption{Schematic force-flux representation of the complete thermodynamic system: a) general configuration; b) simplified configuration for low duration efforts.} 
\label{figure1}
\end{figure}

We now turn to the constitutive equations describing an animal's overall energy balance considering an assembly of $N$ separated (identical) muscle units connected in parallel, and contributing to the production of the total mechanical power. On this large scale, the total incoming and outgoing energy fluxes are $\Phi_{+}=N\varphi_+$ and $\Phi_{-}=N\varphi_-$, and the resistances are $R_E=\frac{r_E}{N}$, $R_M=\frac{r_M}{N}$, $R_+=\frac{r_+}{N}$ and $R_-=\frac{r_-}{N}$. The forces produced by these parallel elements add up, as well as the total metabolic intensity, $I_M= N i_M$, which characterizes the effort produced by the animal. The power and potentials thus satisfy \cite{Goupil2019}: 
\begin{eqnarray}
\Phi_{+} &=& N \varphi_+ = \alpha \mu_{M+} I_{M} + \Delta\mu_M/R_E \\
\Phi_{-} &=& N \varphi_- =\alpha \mu_{M-} I_{M} + R_{M}I_M^2 + \Delta\mu_M/R_E\\
P_{M}    &=& N p_M=\Phi_{+} -\Phi_{-} 
\end{eqnarray}
\noindent where $\Delta\mu_M=\mu_{M+}-\mu_{M-}$ is the chemical potential difference across the conversion zone, and $\alpha$ is the strength of the energy-matter coupling characterizing also the energy conversion efficiency. Since we assume efforts of limited duration, waste production is small and its rejection to the sink not hindered by its accumulation; hence we may consider the limit $r_- \rightarrow 0$ without loss of generality. Note that the zero intensity configuration $I_M = 0$ corresponds to the situation with an organism at rest and a \emph{nonzero} basal residual energy consumption $B = Nb = \frac{\mu_+ -\mu_-}{R_E + R_ +}$ that sustains basic biochemical processes, so that the whole power $\Phi_{+} =\Phi_{-}$ consumed by the organism is used to keep it alive, without production of any (macroscopic) work. From \cite{Goupil2019}, we obtain the power delivered by a \emph{single} muscle unit as the product of the extensive metabolic intensity $I_M$, and the intensive metabolic force per muscle unit, $F_M $:
\begin{eqnarray}
 p_{M}&=&F_{M}i_{M}=\left[F_{\rm iso}-\left(1+\frac{r_{H}}{r_M}\right) r_{M}i_{M}\right]i_{M}
 \label{eq:singleP_M}
\end{eqnarray}
\noindent where $F_{\rm iso}$ is the isometric force for a given muscle unit. Note the presence of the additional dissipative term $R_{H}=\frac{{F_{\rm iso}}+R_{\mathrm{fb}}I_{T}}{I_{T}+I_{M}}$ in Eq.~\eqref{eq:singleP_M}, which stems from feedback effects ${R_{\mathrm{fb}} =\frac{\alpha\mu_{M-}}{I_{T}}}$ \cite{Goupil2019}; the term $I_{T}={N}i_{T}=\frac{1}{\alpha}\frac{R_{E}+R_{+}}{R_{+}R_{E}}$ refers to a threshold of metabolic intensity beyond which the available power collapses. In the case of a Dirichlet-type coupling with the reservoirs, i.e. $R_+=R_-=0$, $R_{H}(I_{M})=0$, there is no feedback effect. As such, the metabolic intensity characterizes the operating point
of the system, i.e., the intensity of the effort produced, either in a static situation or when setting in motion. 

\subsection{Experimental conditions of Hoyt's and Taylor's work}
\label{axe:Hoyt}
It is customary, as Hoyt and Taylor did in \cite{Hoyt1981}, to tilt the treadmill slightly to prevent the subject from working without any effort, which is a situation experienced as unpleasant for the limbs. At constant speed, this experimental treadmill configuration corresponds to a constant average resistive force stress $F_r = f$ exerted on the animal. From an experimental point of view Hoyt and Taylor state that no change in blood lactate levels was detected in the animal up to speeds of 10 m$\cdot$s$^{-1}$ \cite{Hoyt1981}. It is therefore reasonable to consider that no significant anaerobic contribution is to be expected in these measurements, which places them within the limits of validity for the model. The horse is placed on a treadmill, the running speed of which is imposed. It is therefore immobile in relation to the laboratory frame of reference, and even at high speed, there is no external viscous contribution to the force deployed by the animal to move. It can be noted (see Fig.~2 in the main text of the article) that the $COT$ seems to show a discrepancy, and a deviation from the model, for the highest values of speed in the case of walking. This can be understood considering that the animal is in this case in close proximity to the maximum power it can produce when walking, and therefore to its maximum speed for this mode of movement. The linear approximation $I_{M}=kv$ described in Eq.~\eqref{eq_i_de_v} and Fig.~\ref{I_de_v} is then no longer valid, which explains why the points no longer follow the model. Concretely, the animal is in pain, as a walker would be during an exaggeratedly fast walk. This is a physiological state beyond the scope of the model, so we have chosen not to include these points when processing the data.

In their study, the authors did a linear fit to model the relationship between oxygen consumption and velocity. This approach allows to reasonably recover the observed monotonic growth with a relatively small curvature. However, this purely descriptive approach amounts to neglecting dissipation characterized by the quadratic term in Eq.~\eqref{eq:PHI-} and $R_M=0$. In this ideal situation, there is no longer any limitation on the mechanical power growth (see Eq.\,\eqref{eq:PM} of the Supplemental Material and Figs.~1-c--d) and the animal could in principle reach any velocity. Simultaneously the efficiency tends towards a ``Carnot-like'' efficiency $\eta = 1 - \frac{\mu_{M-}}{\mu_{M+}}$. In this case the $COT$ becomes a monotonously decreasing function of the velocity, see Eq.\,\eqref{eq:cot*}. The COT minimum, $COT^*$, therefore diverges and cannot be used to define an optimum velocity as stated by the authors.

\subsection{Relationship between the number of muscle units $N$ and the metabolic intensity $i_M$}
\label{axe:cothill}
In the most general case where only the required power is imposed, neither $N$ nor $i_M$ are fixed a priori, and any increase in the power and/or speed setpoint results in an increase of both $N$ and $i_M$; $N$ can thus vary from $N_0=N(i_M=0)$ to $N_H=N(i_M=I_H)$, i.e. the maximum metabolic intensity explored when all the muscle units are activated. The latter case corresponds to the protocol for an isolated muscle in which all fibers are activated simultaneously, typically force/speed experiments. As a first approximation, let us consider that the relation linking $N$ to $i_M$ can be approximated by a first-order polynomial in $i_M$ for values between $i_M=0$ and $i_M=i_H$,

\begin{eqnarray}
N(i_M) &=& N_0 \left[\left(\frac{N_H}{N_0} -1 \right)\frac{i_M}{i_H} +1 \right]
\end{eqnarray}

\noindent Rewriting the expression of the waste rejection flux $\Phi_-$ as follows, makes its dependency on $N$ appear: 

\begin{eqnarray}
\Phi_- &=& N(a_0 i_M +r_M i^2_M +b) \\
&=& a_0 I_M + \frac{N_H}{N}R_M I^2_M +\frac{N}{N_H}B, 
\end{eqnarray}   
with $R_M= r_M / N_H$, $B= N_H b$ and $I_M = N i_M $. The power is then written 
\begin{eqnarray}\label{eq:PM}
P_M & = &  \left( \alpha \Delta \mu_M - R_M I_M \right) I_M
\end{eqnarray}

\noindent As expected the driving force term $\alpha \Delta \mu_M - R_M I_M$ is intensive, i.e. it does not depend on $N$. Increasing the power $P_M$ can be obtained by multiplying the number of muscle fibers by or increasing the metabolic intensity (up to a certain point).

\begin{figure}[h]
	\includegraphics[width=0.45\textwidth]{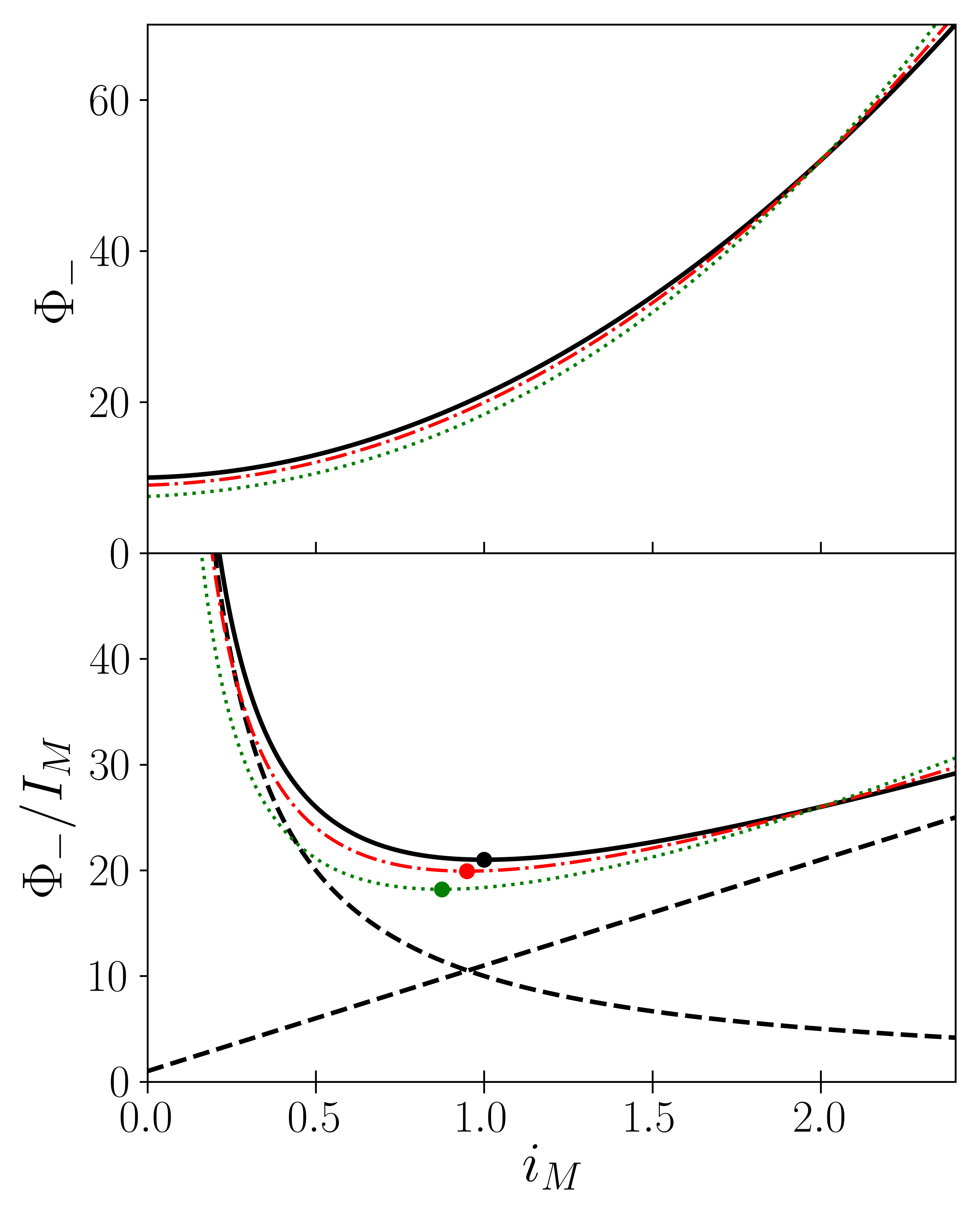}
	\caption{In the top panel $\Phi_-$, represented by the solid thick line, is shown as a function of $i_M \propto v$, for a constant muscle units number $N$; it is also shown as a red dotted-dashed line and a green dotted line for $N$ increasing linearly from $N(i_M=0)=0.90N_H$ to $N(i_M=I_H)=N_H$, and from $N(i_M=0)=0.75N_H$ to $ N(i_M=I_H)=N_H$ with $I_H=2$, respectively. In the bottom panel, the corresponding $COT=\Phi_-/v$ is shown as a function of $i_M\propto v$ (with the same same color code). The dots indicate the minimum of each curve. The ratio $B/v$ is represented by the decreasing black dashed curve, while the increasing black dashed line that shows $a_0+R i_M$ (see Eq.~(5) in the main text) characterizes the dissipation process.}
	\label{I_de_v}
\end{figure}

Considering a muscle unit that contracts at the frequency $f$ over a length $L$, the metabolic intensity $i_M$ can be approximated using a linear relationship $i_M \propto f L = k v_1$, with $v_1$ being the global velocity obtained using a single muscle unit and $k$ a coupling constant. Thus, the speed $v$ associated with $N$ fibers is written $I_M = N i_M = N k v_1 = k v $, where $v$ is the observed velocity (the horse forward motion). For an animal moving on an inclined slope, modulating the angle of this slope while keeping constant $v$ is equivalent to increase $N$ keeping $f$ constant. Conversely, modulating $f$ at constant $N$ is equivalent to a classical force/speed experiment. For any displacement the metabolic power $P_M$ and the external required power are related via $P_r= F_r v$, where $F_r$ is the required force and $P_r \leq P_M$. Of course $F_r$ depends directly on the experimental conditions, i.e. the viscous friction due to the air, the slope, the transported load. In the case of horizontal movement on a conveyor belt, it is reasonable to assume that $F_r$ is  constant. Thus it comes from Eq.~(4) in the main text that the general expression for $v(I_{M})=\frac{F_M I_M}{F_r} $ may read:

\begin{equation}
v = \left[F_{\rm iso} - \left(1+\frac{R_{H}(I_M)I_{M}}{R_M} \right) R_{M}I_{M}\right] \frac{I_{M}}{F_r} 
\label{eq_i_de_v}
\end{equation}

\noindent from which we then derive the $COT$: 

\begin{eqnarray}
\textit{COT}    &=& \frac{N}{N_H} \left( a_0 k + R_M k^2 v + \frac{B}{v}\right) \\
                &=& k \frac{N}{N_H} COE_-  
\label{eq:COTa}
\end{eqnarray}

\noindent thus introducing the energy cost of effort $COE_-$.
Contrary to the situation when all the muscle units are stimulated, both the effective basal and effective viscosity depend on the operating point. The basal is modulated downwards by a factor $\frac{N}{N_H}<1$ while the effective viscosity is increased by the inverse of this factor, $\frac{N_H}{N}>1$. When $N=N_H$ is constant, it comes that the speed associated with the minimum of $COT$ reads:

\begin{equation}
v^*=\sqrt{\frac{r_M b}{i_M}}
\end{equation}

\noindent which, in a scaled version, can be written:

\begin{equation}
V =  \frac{R_M I_T^2}{F_r v^*} \frac{I_M^2}{I_M+1} \left[\left( \frac{F_{\rm iso}}{a_0} \frac{1}{I_M} + 1 \right) r z - I_M + 1\right]
\label{eq_i_i_de_v_ad}
\end{equation}

\noindent where $I=i_M/I_T$ and $V=v/v^*$ indicate the scale (dimensionless) for intensities and velocities respectively; $z$ is the figure of merit of the underlying thermodynamic process, which is a generalization of the figure of merit usually encountered in, e.g., thermoelectricity: $z=\frac{F_{\rm iso}R_{\rm fb}I_T}{R_MB}$. Note that in the case of a strict Dirichlet type boundary condition, i.e. $R_+=0$, the above equation is reduced to
  
\begin{equation}\label{eq:V}
V =  \frac{R_M I_T^2 }{f v^*}I_M^2
\end{equation}

\subsection{Remarks on the energy cost of effort $COE_-$}
The cost of effort, $COE_{-}$, introduced in Eq.~\eqref{eq:COTa}, is based on a few simple elements that define a thermodynamic conversion system. The first element that underpins the existence of a $COE_{-}$ is the presence of boundary conditions that lie somewhere between Dirichlet and Neumann, i.e. that $R_+$ and/or $R_-$ are not zero. The second element is the basal power, which itself relies on the existence of a bridging element with finite resistance, $R_{E}$, which acts as a bypass of the energy flux within the conversion element, i.e. a leakage of part of the energy between the two reservoirs, even in the absence of any power production, i.e. even for $I_M=0$. Note that the presence of a basal power does not condition the existence of $COE_{-}$, but more precisely the existence of a minimum for $COE_{-}^*$, which corresponds to the minimum waste production.

The flux $\Phi_{-}$ and the corresponding $COE_-=\Phi_{-}/I_M = COE_- = N_HCOT/kN$ are shown against $i_M$ (from $i_M=0$ to $i_M=i_H$) in Fig.~\ref{I_de_v} with the number of muscle units varying by 10\% for one case, and 25\% for the other case. As expected, the overall behavior is preserved and, in particular, the hyperbolic behavior when $i_M \rightarrow 0$ as well as the linear growth beyond the $COE_-$ minimum. Because the basal $B=\frac{N}{N_H}B$ reduces as the dependence of $N$ in $i_M$ increases, it is also expected that the coordinates of the minimum $(v^*,COE_-^*)$ shift towards lower values. It is interesting to observe that at constant $N$ the COT reduces to $ COT \approx a_0 k + R_M k^2 v $ for high velocities and, consequently, that the term $ a_0 k $ appears as the intercept in $v = 0$ of this reduced form. In the case discussed in this section where $N$ grows with the metabolic intensity, the term $a _v = \frac{N}{N_H}a_0 $, which is extracted from the experimental data, is such that $a_v< a_0$, as it appears clearly in Fig.~\ref{I_de_v}. Thus, if $a_0 = \alpha \mu_-$ is small, $a_v$ can be observed negative, as reported in table\,\ref{tab:bilan_thermo}. This question remains unresolved, due to the lack of Hill-type measurements that would remove the uncertainty on the determination of $a_0$.

\end{widetext}

\end{document}